\documentstyle[aps%,twocolumn
]{revtex}
\begin{document}
\draft
\title{Quantum Critical Behavior of Disordered Superfluids }
\author{Adriaan M. J. Schakel}
\address{Institut f\"ur Theoretische Physik \\ Freie Universit\"at Berlin \\
Arnimallee 14, 14195 Berlin}
\date{April 7, 1996}
\maketitle
\begin{abstract}
The quantum critical behavior of an interacting, non-relativistic Bose theory
with quenched disorder randomly distributed in space is investigated.  The
renormalization group is carried out in a double $\epsilon$ expansion, where one
$\epsilon$ is the deviation of the effective space-time dimensionality from 4,
while the other denotes the number of time dimensions.  The disordered theory,
which displays localization in the superfluid state, is shown to possess an
infrared stable fixed point.
\end{abstract}

Recently, it was demonstrated that in the absence of disorder the
Chern-Simons-Ginzburg-Landau (CSGL) model has a second-order quantum phase
transition \cite{AS}.  The model, capturing the essential properties of a
quantum Hall liquid, gives an effective description of the fractional quantized
Hall effect (FQHE) \cite{Zhang}.  The model is dubbed CSGL because it is on a
similar footing as the Ginzburg-Landau model of superconductivity, and because
it involves a statistical gauge field governed by a Chern-Simons term.  This
gauge field accounts for the composite particles featuring in the FQHE.  Since
disorder plays an essential part in the FQHE, a realistic description of the
critical behavior of a quantum Hall liquid should include impurities.

The CSGL model is based on a non-relativistic $(\phi^* \phi)^2$ theory.
The quantum critical behavior of this theory was investigated
with help of the renormalization group in Ref.\ \cite{Uzunov}.  Below the upper
critical (space) dimension $d_{\rm c} =2$---the dimension of interest to us, the
theory was shown to possess a non-trivial infrared stable (IR) fixed point,
implying that the system undergoes a second-order phase transition.  Because of
the analytic structure of the non-relativistic propagator, the critical
exponents characterizing the transition are not affected by quantum fluctuations
and retain their Gaussian values.

More than 10 years ago, it has been argued in these pages \cite{KU} that upon
including quenched disorder, the quantum critical behavior of the
non-relativistic $(\phi^* \phi)^2$ theory becomes unstable.  Only after
introducing an artificial high-frequency cutoff into the problem, an IR fixed
point was found by Weichman and Kim \cite{WK}.  However, as was pointed out by
the authors, such a cutoff is difficult to justify as it would imply that time
is discrete.  So, it is widely accepted among the researches in the field that
the disordered non-relativistic $(\phi^* \phi)^2$ theory has no perturbatively
accessible IR fixed point, if any at all \cite{FWGF}.  The absence of an IR
fixed point in the non-relativistic $(\phi^* \phi)^2$ theory would imply that
also the quantum critical behavior of the CSGL model, being based on this
theory, is unstable with respect to impurity influences.  It would then be hard
to envisage a continuous phase transition for the FQHE of the type discussed in,
for example, Ref.\ \cite{KLZ}.

Because of its implications for the description of the critical behavior of
the FQHE we have revisited the problem.  Below, it will be shown that, contrary
to general conviction, the disordered non-relativistic $(\phi^* \phi)^2$
theory has a new IR fixed point.  The calculations are performed without
introducing a (physically unnatural) high-frequency cutoff.  

The non-relativistic $(\phi^* \phi)^2$ theory at finite density is described by
the Lagrange density
\begin{equation}
\label{Lneutral}
{\cal L} = \frac{i}{2} \phi^* \tensor{\partial}_t \phi - \frac{1}{2} |\nabla
\phi|^2 + [\mu + \psi({\bf x})] \phi^* \phi - \frac{u }{4}(\phi^* \phi)^2,
\end{equation}
with $\tensor{\partial}_t = \partial_t - \loarrow{\partial_t}$ the time
derivative operating to the right and left, $\mu$ the chemical potential,
$\psi({\bf x})$ the random field and $u>0$ the strength of a $\delta$-function
repulsion.  We have rescaled the fields and parameters so that the mass
parameter disappeared from the theory.  In the limit $\mu \rightarrow 0$, the
theory becomes critical.  The theory will be studied in the symmetrical state
where the chemical potential is negative and the global U(1) symmetry unbroken.
We therefore set $\mu = - r/2$, with $r>0$.

The random field $\psi({\bf x})$ is assumed to be Gaussian distributed
\cite{Ma}:
\begin{equation} 
P(\psi) = \exp \left[-\frac{1}{\Delta} \int {\rm d}^d x \, \psi^2({\bf x})
\right],
\end{equation}
characterized by the parameter $\Delta$.  Since $\psi({\bf x})$ depends only on
the $d$ spatial dimensions, the impurities it describes should be considered as
grains randomly distributed in space.  When time is included, as is required for
the study of quantum critical phenomena, the static grains trace out straight
worldlines.  That is to say, the impurities are line-like.

It has been shown by Dorogovtsev \cite{Dorogovtsev} that the critical
properties of systems with extended defects must be studied in a double
$\epsilon$ expansion, otherwise no IR fixed point is found.  The method
differs from the usual $\epsilon$ expansion, in that it also includes an
expansion in the defect dimensionality $\epsilon_{\rm d}$.  To carry out this
program in the present context, where the defect dimensionality is determined
by the dimensionality of time, the theory has to be formulated in
$\epsilon_{\rm d}$ time dimensions.  The case of interest is $\epsilon_{\rm
d}=1$, while in the opposite limit, $\epsilon_{\rm d}\rightarrow 0$, the
disordered non-relativistic $(\phi^* \phi)^2$ reduces to the classical spin
model with random (point-like) disorder.  Hence, $\epsilon_{\rm d}$ is a
parameter with which quantum fluctuations can be suppressed.  An expansion in
$\epsilon_{\rm d}$ is a way to perturbatively include the effect of quantum
fluctuations on the critical behavior.  Ultimately, we will be interested in
the case $\epsilon_{\rm d}=1$.

The zero-temperature partition function $Z$ of the disordered system in
$\epsilon_{\rm d}$ time and $d$ spatial dimensions is given by the functional
integral
\begin{equation} \label{Z}
Z = \int D \psi \, D \phi^* D \phi \, P(\psi)  \exp\left(i^{\epsilon_{\rm d}} \int
{\rm d}^{\epsilon_{\rm d}} t \, {\rm d}^d x \, {\cal L} \right),
\end{equation} 
where ${\cal L}$ is the Lagrange density (\ref{Lneutral}) generalized to
$\epsilon_{\rm d}$ time dimensions.  It involves instead of just one time
derivative, a sum of $\epsilon_{\rm d}$ derivatives: $ 
\partial_t \rightarrow \partial_{t_1} + \partial_{t_2} + \cdots +
\partial_{t_{\epsilon_{\rm d}}}.   
$
The factor $i^{\epsilon_{\rm d}}$ appearing in (\ref{Z}) arises from the
presence of $\epsilon_{\rm d}$ time dimensions, each of which is accompanied
by a factor of $i$.  The integral over the random field is easily carried out
to yield for the partition function
\begin{equation} 
Z = \int D \phi^* D \phi  \exp \left[i^{\epsilon_{\rm d}} \int
{\rm d}^{\epsilon_{\rm d}} t \, {\rm d}^d x \, {\cal L} + i^{2
\epsilon_{\rm d}} \frac{\Delta}{4}  \int
{\rm d}^{\epsilon_{\rm d}} t \, {\rm d}^{\epsilon_{\rm d}} t' \, {\rm d}^d x \,
|\phi(t,{\bf x})|^2 |\phi(t',{\bf x})|^2  \right]. 
\end{equation} 
The randomness is seen to result in a quartic interaction term which is
non-local in time.  To calculate the quantum critical properties of this
disordered theory we will not employ the replica method \cite{GL}, but instead
follow Lubensky \cite{Lubensky}.  In this approach, the averaging over
impurities is carried out for each Feynman diagram separately.  The upshot is
that only those diagrams must be included which remain connected when $\Delta$,
the parameter characterizing the Gaussian distribution of the impurities, is set
to zero \cite{Hertz}.

Following Weichman and Kim \cite{WK}, we evaluate the integrals over loop
frequencies assuming that all frequencies are either positive or negative.  This
allows us to employ Schwinger's proper-time representation of propagators
\cite{Schwinger}, which is based on the integral representation of the $\Gamma$
function,
\begin{equation} 
\frac{1}{a^z} = \frac{1}{\Gamma(z)} \int_0^\infty \frac{{\rm d} \tau}{\tau}
\tau^z {\rm e}^{-\tau a}.
\end{equation}  
The frequency integrals we encounter to the one-loop order can be carried out
with the help of the equations
\begin{mathletters} \label{int} 
\begin{eqnarray}  
\int' \frac{{\rm d}^{\epsilon_{\rm d}} \omega}{(2\pi)^{\epsilon_{\rm d}}}
\frac{1}{\omega_1 + \omega_2 + \cdots + \omega_{\epsilon_{\rm d}} -x
\pm i \eta} &=& -\frac{\Gamma(1-\epsilon_{\rm d})}{(2\pi)^{\epsilon_{\rm d}}}
{\rm sgn}(x) |x|^{\epsilon_{\rm d}-1} \left({\rm e}^{\pm i \, {\rm sgn}(x) \pi
\epsilon_{\rm d}} + 1 \right), \label{inta} \\ 
\int' \frac{{\rm d}^{\epsilon_{\rm d}} \omega}{(2\pi)^{\epsilon_{\rm d}}}
\frac{{\rm e}^{i(\omega_1 + \omega_2 + \cdots + \omega_{\epsilon_{\rm
d}})\eta}}{\omega_1 + \omega_2 + \cdots + \omega_{\epsilon_{\rm 
d}} -x 
+ i x \eta} &=&  \frac{i \pi}{(2\pi)^{\epsilon_{\rm d}}\Gamma(\epsilon_{\rm
d})} 
(i|x|)^{\epsilon_{\rm d}-1} \left[ \sin(\case{1}{2} \pi \epsilon_{\rm d}) -
\frac{{\rm sgn}(x)}{\sin(\case{1}{2} \pi \epsilon_{\rm d} )} \right],
\label{intb} 
\end{eqnarray}  
\end{mathletters}
where $\eta$ is an infinitesimal positive constant which is to be taken to
zero after the frequency integrals have been carried out.  The prime on the
integrals is to remind the reader that the frequency integrals are taken over
only two domains with either all frequencies positive or negative.  By
differentiation with respect to $x$, Eq.\ (\ref{inta}) can, for example, be
employed to calculate integrals involving integrands of the form $1/(\omega_1
+ \omega_2 + \cdots + \omega_{\epsilon_{\rm d}} -x + i \eta)^2$.  Is is easily
checked that in the limit $\epsilon_{\rm d} \rightarrow 1$, where the
frequency integral can be performed with help of contour integration, Eqs.\
(\ref{int}) reproduce the right results.  When considering the limit of zero
time dimensions ($\epsilon_{\rm d} \rightarrow 0$), it should be remembered
that the frequency integrals were taken over two separate domains with all
frequencies either positive or negative.  Each of these domains is contracted
to a single point in the limit $\epsilon_{\rm d} \rightarrow 0$, so that one
obtains a result which is twice that obtained by simply purging any reference
to the time dimensions.  The integral (\ref{intb}) has an additional
convergence factor $\exp(i\omega \eta)$ for each of the $\epsilon_{\rm d}$
frequency integrals.  This factor, which is typical for non-relativistic
quantum theories \cite{Mat}, is to be included in self-energy diagrams
containing only one propagator.

Before studying the disordered theory, let us briefly consider the
non-relativistic $(\phi^* \phi)^2$ theory in the absence of impurities.  In this
case, there is no need for an $\epsilon_{\rm d}$ expansion and the formalism
outlined above should yield results for arbitrary $0 \leq \epsilon_{\rm d} \leq
1$, which interpolate between the classical and quantum limit.  After the
frequency integrals have been performed with the help of Eqs.\ (\ref{int}), the
standard technique of integrating out a momentum shell can be applied to obtain
the renormalization group equations.  For the correlation-length exponent $\nu$
we obtain in this way
\begin{equation} \label{nupure}
\nu = \frac{1}{2} \left[1 + \frac{\epsilon}{2} \frac{m+1}{(m+4) - (m+3)
\epsilon_{\rm d}} \cos^2( \case{1}{2} \pi \epsilon_{\rm d}) \right].
\end{equation}   
Here, $\epsilon = 4-2\epsilon_{\rm d}-d$ is the deviation of the {\it effective}
space-time dimensionality from 4, where it should be noted that in (canonical)
non-relativistic theories, time dimensions have an engineering dimension twice
that of space dimensions.  (This property is brought out by the
Gaussian value $z=2$ for the dynamical exponent $z$.)  For comparison we have
extended the theory (\ref{Lneutral}) to include $m$ complex $\phi$ fields
instead of one field.  In the classical limit, Eq.\ (\ref{nupure}) gives the
well-known one-loop result for a classical spin model with $2m$ real components,
$\nu \rightarrow \frac{1}{2} [1 + \frac{1}{2} \epsilon (m+1)/(m+4)]$, while in
the quantum limit it gives the result $\nu \rightarrow \frac{1}{2}$, as required
\cite{Uzunov}.  The exponent (and also the fixed point) diverges when
$\epsilon_{\rm d} \rightarrow (m+4)/(m+3)$.  Since this value is always larger
than one, the singularity is outside the physical domain $0 \leq \epsilon_{\rm
d} \leq 1$.  This simple example illustrates the viability of the formalism
developed here to generate results interpolating between the classical and
quantum limit.

Let us continue with the disordered theory.  After the frequency integrals
have been carried out, it is again straightforward to derive the
renormalization group equations by integrating out a momentum shell
$\Lambda/b<k<\Lambda$, where $\Lambda$ is a high-momentum cutoff and
$b=\exp(l)$, with $l$ infinitesimal.  Defining the dimensionless variables
\begin{equation} 
\tilde{u} = \frac{K_d}{(4 \pi)^{\epsilon_{\rm d}}} u
\Lambda^{-\epsilon}; \;\;\;
\tilde{\Delta} = K_d \Delta \Lambda^{d-4}; \;\;\;
\tilde{r} = r \Lambda^{-2},
\end{equation} 
where $K_d= 2/(4\pi)^{d/2}\Gamma(d/2)$ is the area of a unit sphere in $d$
spatial dimensions divided by $(2\pi)^d$, we find
\begin{mathletters} 
\begin{eqnarray} \label{reneq} 
\frac{{\rm d} \tilde{u}}{{\rm d} l} &=&  \epsilon \tilde{u}  -8
\left[\Gamma(1-\epsilon_{\rm d}) + (m+3) \Gamma(2-\epsilon_{\rm d}) \right]
\cos(\case{1}{2}\pi \epsilon_{\rm d}) 
\tilde{u}^2 + 24 \tilde{\Delta} \tilde{u} \label{reneqa} \\
\frac{{\rm d} \tilde{\Delta}}{{\rm d} l} &=&  (\epsilon + 2\epsilon_{\rm
d})\tilde{\Delta }  + 16 \tilde{\Delta}^2 - 16 (m+1)
\Gamma(2-\epsilon_{\rm d}) \cos(\case{1}{2}\pi \epsilon_{\rm d} ) 
\tilde{u} \tilde{\Delta} \label{reneqb} \\
\frac{{\rm d} \tilde{r}}{{\rm d} l} &=& 2 \tilde{r} + \frac{4 \pi (m+1)}{
\Gamma(\epsilon_{\rm d})} \frac{\cos^2(\case{1}{2} \pi \epsilon_{\rm d})}
{\sin(\case{1}{2} \pi \epsilon_{\rm d} )}  \tilde{u} -
4 \tilde{\Delta}. \label{reneqc}
\end{eqnarray} 
\end{mathletters}
These results are to be trusted only for small values of $\epsilon_{\rm d}$.
For illustrative purposes we have, however, kept the full $\epsilon_{\rm d}$
dependence.  The set of equations yields the fixed point
\begin{mathletters}
\begin{eqnarray} \label{fp}
\tilde{u}^* &=& \frac{1}{16 \cos(\case{1}{2}\pi \epsilon_{\rm d} )
\Gamma(1-\epsilon_{\rm d})} \, \frac{6 \epsilon_{\rm d} + 
\epsilon}{2m(1-\epsilon_{\rm d}) -1}  \\
\tilde{\Delta}^* &=& \frac{1}{16} \frac{
m(1-\epsilon_{\rm d}) (2 \epsilon_{\rm d} -\epsilon) + 2 \epsilon_{\rm d}
(4-3\epsilon_{\rm d}) + \epsilon (2 -\epsilon_{\rm d})}{2m(1-\epsilon_{\rm d})
-1},
\end{eqnarray}   
\end{mathletters}
and the critical exponent 
\begin{equation} \label{nufull} 
\nu = \frac{1}{2} + \frac{\epsilon +2 \epsilon_{\rm d}}{16} + \frac{(m+1)
(6\epsilon_{\rm d} + \epsilon )  [\epsilon_{\rm d}+\cos( \pi
\epsilon_{\rm d})]}{16[2m(1-\epsilon_{\rm d})-1]}.
\end{equation} 
We see that both $\tilde{u}^*$ and $\tilde{\Delta}^*$ diverge when
$\epsilon_{\rm d} \rightarrow 1-1/2m$.  At this point, the fixed point becomes
unphysical.  The singularity separates the quantum regime $\epsilon_{\rm
d}\lesssim 1$ from the classical regime $\epsilon_{\rm d} \gtrsim 0$ about
which perturbation theory is to be carried out.  When the equations are
expanded to first order in $\epsilon_{\rm d}$, we recover the IR fixed point
found by Weichman and Kim \cite{WK} using an high-frequency cutoff:
\begin{equation} 
\tilde{u}^*= \frac{1}{16} \frac{\epsilon + 6 \epsilon_{\rm d}}{2m-1}; \;\;\; 
\tilde{\Delta}^*= \frac{1}{16} \frac{(2-m)\epsilon + 2(m+4) \epsilon_{\rm
d}}{2m-1}, 
\end{equation}
with the critical exponent
\begin{equation} \label{nuqm}
\nu = \frac{1}{2} \left(1 + \frac{1}{8} \frac{3m \epsilon + (5m +2)
2\epsilon_{\rm d}}{2m-1} \right).
\end{equation} 
We thereby provide support for the existence of this fixed point.  

The value of the critical exponent (\ref{nuqm}) should be compared with that
of the classical spin model with $2m$ components in the presence of random
impurities of dimension $\epsilon_{\rm d}$ \cite{Dorogovtsev}:
\begin{equation} 
\nu = \frac{1}{2} \left(1 + \frac{1}{8} \frac{3m \epsilon + (5m +2)
\epsilon_{\rm d}}{2m-1} \right).
\end{equation}  
Taking into account that in a non-relativistic quantum theory, time dimensions
count double as compared to space dimensions, we see that both results are
equivalent. 

The limit of interest to us, corresponding to $m=1$, $\epsilon=0$, and
$\epsilon_{\rm d} =1$, is probably difficult to reach by low-order
perturbation theory, for the quantum regime is separated by a singularity from
the classical regime where perturbation theory applies.  Although this might be
an artifact of the one-loop calculation, it is unlikely that by including a
few more loops, the quantum regime becomes accessible via the classical
regime.  Because the singularity moves towards $\epsilon_{\rm d}=1$ when the
number of field components increases, one might be tempted to carry out a
$1/m$ expansion.  However, observe that the $m$-dependence in (\ref{fp}) and
(\ref{nufull}) disappears in the quantum limit $\epsilon_{\rm d} \rightarrow
1$.  To understand this property we note that the $m$-dependence stems from
oriented ring diagrams.  Because of the analytic structure of the
non-relativistic propagator, these diagrams and therefore all $m$-dependences
vanish in the quantum limit \cite{Uzunov}.  A $1/m$ expansion has,
consequently, little bearing on the problem.

If the IR fixed point is to be of relevance to the FQHE, the impurities have to
lead to localization---at least in the superfluid state.  For the model at
hand, albeit in $d=3$, this connection has been established by Huang and Meng
\cite{HM}.  They showed that in this state, the impurities give rise
to an additional depletion of the condensate as well as of the superfluid
density.  (The latter is defined by considering the response of the system to an
externally imposed velocity field as specified by the expression for the
momentum density ${\bf j}$.)  They found that the depletion of the superfluid
density is larger than that of the condensate.  Apparently, part of the
zero-momentum states belongs to the normal fluid rather than to the condensate.
They interpreted this as indicating that these states are trapped by impurities.

This situation should be contrasted to the one in the absence of impurities.  In
this case, the condensate is depleted only due to the inter-particle repulsion.
Despite the depletion, {\it all} the particles participate in the superflow
motion at zero temperature, the momentum density being given by ${\bf j} = m n
{\bf u}_{\rm s}$.  Here, $n$ is the {\it total} particle density, ${\bf u}_{\rm
s}$ denotes the velocity with which the condensate moves, i.e., the superfluid
velocity, and we have reintroduced the mass parameter $m$.  This equation leads
to the conclusion that the normal fluid is dragged along by the condensate.

In $d$ spatial dimensions, we find for the depletion $n_\Delta$ of the
condensate due to impurities:
\begin{equation} 
n_\Delta = \frac{2^{d/2-5}\Gamma(2-d/2)}{\pi^{d/2}} m^{d/2} u^{d/2-2}
n_0^{d/2-1} \Delta,
\end{equation}   
where $n_0$ denotes the density of particles residing in the condensate.  Note
that in the case of interest to us, the depletion is independent of
the condensate density $n_0$:
\begin{equation} 
n_\Delta = \frac{1}{16 \pi} \frac{m}{u} \Delta \;\;\;\;\; {\rm for} \;\;\; d=2.
\end{equation}   
For the  momentum density ${\bf j}$ we obtain
\begin{equation} 
{\bf j} = m n {\bf u}_{\rm s} + \frac{4}{d} m n_\Delta ({\bf u}- {\bf 
u}_{\rm s} ),
\end{equation} 
with ${\bf u}$ denoting the externally imposed velocity.  The particle density
moving with the velocity ${\bf u}$ belongs to the normal fluid.  We see that it
is a factor $4/d$ larger than the depletion of the condensate $n_\Delta$ due to
impurities.  (For $d=3$ this gives the factor $\case{4}{3}$ first found in Ref.\
\cite{HM}.)  In other words, for every space dimension smaller than four, part
of the zero-momentum states belongs to the normal fluid, and is trapped by
impurities.  Upon invoking a statistical gauge field, this superfluid state
represents a quantum Hall liquid.  Hence, the phenomenon of localization, which
is essential for understanding the FQHE, can be accounted for in the CSGL model
by including (quenched) disorder.

In principle this is a good starting point to study the quantum critical
behavior of the CSGL model in the presence of impurities, all the prerequisites
(existence of an IR fixed point in the neutral model, localization) being
fulfilled.  There is, however, a technical problem.  We have worked in a double
$\epsilon$ expansion thereby leaving physical space-time.  Unfortunately, the
Chern-Simons term governing the statistical gauge field of the model is defined
only in $2+1$ dimensions.  This makes a study of the critical behavior of the
disordered CSGL model in a double $\epsilon$ expansion impossible, and a
different method, probably non-perturbative in character, is needed.

\end{document}